\documentclass{article}
\usepackage[T1]{fontenc} 
\usepackage[utf8]{inputenc} 
\usepackage{ismir,amsmath,cite,url}
\usepackage{graphicx}
\usepackage{color}
\usepackage{pifont}
\usepackage{array}
\usepackage{multirow}
\newcommand{\cmark}{\ding{51}}%
\newcommand{\xmark}{\ding{55}}%

\usepackage{microtype}
\usepackage{paralist}
\usepackage{units}
\usepackage[bookmarks=false,hidelinks]{hyperref}

\usepackage{lineno}

\title{Multitask learning for instrument activation aware music source separation }
\twoauthors
{Yun-Ning Hung} {Center for Music Technology\\
  Georgia Institute of Technology\\
{\tt amyhung@gatech.edu}}
{Alexander Lerch} {Center for Music Technology\\
  Georgia Institute of Technology\\
{\tt alexander.lerch@gatech.edu}}

\begin{document}

\maketitle
\begin{abstract}
Music source separation is a core task in music information retrieval which has seen a dramatic improvement in the past years. Nevertheless, most of the existing systems focus exclusively on the problem of source separation itself and ignore the utilization of other~---possibly related---~MIR tasks which could lead to additional quality gains. In this work, we propose a novel multitask structure to investigate using instrument activation information to improve source separation performance. Furthermore, we investigate our system on six independent instruments, a more realistic scenario than the three instruments included in the widely-used MUSDB dataset, by leveraging a combination of the MedleyDB and Mixing Secrets datasets. The results show that our proposed multitask model outperforms the baseline Open-Unmix model on the mixture of Mixing Secrets and MedleyDB dataset while maintaining comparable performance on the MUSDB dataset.
\end{abstract}

\section{Introduction}
Music source separation has long been an important task for Music Information Retrieval (MIR) with numerous practical applications. By isolating the sound of individual instruments from a mixture of music, source separation systems can be used, for example, as a pre-processing tool for music transcription \cite{paulus2005drum} or for audio remixing \cite{veire2018raw}. They also enable special applications such as the automatic generation of karaoke tracks by separating vocals from the accompaniment, stereo-to-surround upmixing, and instrument-wise equalization \cite{rafii2018overview, adali2014source}.

Most of the current source separation systems use deep learning approaches to estimate a spectral mask for each independent instrument, then apply the mask to the mixture audio for separation. Although the utilization of deep learning has improved source separation performance dramatically, one problem of this approach is the limited amount of training data for the prevalent supervised learning approaches. More specifically, the datasets need to comprise of the separated tracks of each instrument, which renders most easily accessible music data useless as it is already mixed. Multiple open-source datasets attempt to address this issue \cite{SiSEC16, Chan2015VocalAI, Hsu2010OnTI}. MUSDB \cite{musdb18} is nowadays the most frequently used dataset for the training and evaluation of source separation systems. In addition to the limited size, the main shortcoming of MUSDB is the limited number of instrument tracks: `Bass,' `Drums,' `Vocals,' and `Other.' Other datasets show other drawbacks, for instance, the iKala and MIR-1K datasets only contain short clips of music instead of complete songs \cite{Chan2015VocalAI, Hsu2010OnTI}.

In addition to the data challenge, one potential issue with most existing music source separation systems is that they exclusively focus on the source separation task itself. Harnessing the information of other MIR tasks by incorporating them into source separation, however, has not been explored in-depth. For example, Instrument Activation Detection (IAD) can help determine which time frame contains the target instrument, while pitch detection can help determine which frequency bins are more likely to contain a harmonic series \cite{Miron2017MonauralSS, Hung2018FramelevelIR}. This kind of multitask learning approach has been reported to be efficient for multiple other MIR tasks. B{\"o}ck et al.\ achieve state-of-the-art performance for both tempo estimation and beat tracking by learning these two tasks at the same time \cite{Bck2019MultiTaskLO}. Bittner et al.\ show that by estimating multi-f0, melody, bass line, and vocals at the same time, the system outperforms its single-task counterparts on all four tasks \cite{Bittner2018MultitaskLF}. Similar results have been reported for simultaneously estimating score, instrument activation, and multi-f0 \cite{Hung2019MultitaskLF}. However, only recently was multitask learning successfully applied to source separation by combining it with pitch estimation \cite{seetharaman2019class}.

In this paper, we propose a novel multitask learning structure to explore the combination of IAD and music source separation. By training for both tasks in an end-to-end manner, the estimated instrument labels can be used during inference as a weight for each time frame. The goal is both to suppress the frames not containing the target instrument and to correct a potentially incorrectly estimated mask. To increase the size of the available training data, we leverage two open-source large-scale multi-track datasets (MedleyDB \cite{bittner2014medleydb} and Mixing Secrets \cite{gururani2017mixing}) in addition to MUSDB to evaluate on a larger variety of separable instruments. We refer to the combination of these two datasets as the \textit{MM dataset}. 

In summary, the main contributions of this work are
\begin{compactitem}
    \item the systematic investigation of the first multi-task source separation deep learning model that incorporates source separation with IAD in the spectral domain, 
    \item   the application of the IAD predictions during inference, and
    \item the presentation of the first open-source model that separates up to six instruments instead of the four tracks (3 independent instruments) featured by MUSDB. 
\end{compactitem}


\section{Related Work}
State-of-the-art systems for music source separation are all based on deep learning due to proven superior performance. Uhlich et al.\ presented one of the pioneering works using a Deep Neural Network (DNN) architecture for music source separation \cite{Uhlich2015DeepNN}, and Nugraha et al.\ used a DNN architecture and fully-connected layers for multichannel music source separation \cite{nugraha2016multichannel}. In the following years, more deep learning related systems were introduced. For example, Takahashi and Mitsufuji used recurrent neural networks to deal with temporal information \cite{uhlich2017improving}, while others proposed the U-net structure for multiple separation tasks \cite{ jansson2017singing}. The U-net structure had been previously found useful for image segmentation \cite{Ronneberger2015UNetCN} and treats the decomposition of a musical audio signal into the target and accompanying instrument tracks analogous to image-to-image translation. 
Takahashi et al.\ presented a dense LSTM that achieved the highest score in the SiSEC2018  \cite{SiSEC18} competition \cite{takahashi2018mmdenselstm}.  To  preserve high resolution information, Liu and Yang introduced dilated 1-D convolution and a GRU network to replace pooling \cite{liu2019dilated}. Different from other approaches using spectrograms as the input representation, D{\'e}fossez et al.\ experimented on time-domain waveform source separation and showed that results comparable to spectrogram-based source separation systems are achievable \cite{defossez2019music}. ``Spleeter,'' based on a U-net model structure, is currently regarded as one of the most powerful source separation systems \cite{spleeter2019}. It should be noted that~---although the pre-trained model is freely available---~Spleeter is trained on a proprietary, publicly unavailable dataset.  

St\"oter et al.'s ``Open-Unmix'' is frequently used as modern benchmark system on the MUSDB dataset \cite{musdb18}. It is a well documented open-source music source separation system with a recurrent architecture that achieves good separation results \cite{stoter2019open}. 

Most of the methods mentioned above are trained and evaluated on the open-source dataset used in SiSEC2018 competition \cite{SiSEC18}: MUSDB \cite{musdb18}. 
As mentioned above, one of the main problems of the MUSDB dataset is that it has only a limited amount of songs and instrument categories: it only includes three separable independent instruments. To include more separable instruments, Miron et al.\ proposed a score-informed system able to separate four classical instruments by training it on synthetic renditions  \cite{Miron2017MonauralSS}. However, their system is limited to classical music and requires the musical score for separation. While Spleeter is able to separate four independent instruments and Uhlich et al.\ constructed a dataset which contains nine separable instruments \cite{Uhlich2015DeepNN}, both their datasets are not publicly available.

To explore the possibility of separating unseen instruments, Seetharaman et al.\ proposed to use instrument class labels as a condition to cluster time-frequency bins for different instruments into the embedding space \cite{seetharaman2019class}. Their work showed that the system can also separate unseen instruments during testing by sampling from the learned embedding space. Lee et al.\ proposed to use audio queries for music separation \cite{lee2019audio}. The learned feature vector from the audio query acts as the condition to inform the separation of the target source. Manilow et al.\ introduced a multitask learning structure for source separation, instrument classification, and music transcription \cite{manilow2019simultaneous}. They showed that by jointly learning these three tasks, the source separation quality increased. Furthermore, the network seems to generalize better to unseen instruments. Other works such as \cite{slizovskaia2019end, Swaminathan2019ImprovingSV} combine instrument activity with source separation, however, they utilized either the predicted activity or ground truth as an input condition instead of learning in an end-to-end manner.

\begin{figure}[t]
    \centering
    \includegraphics[width =
    0.8\columnwidth]{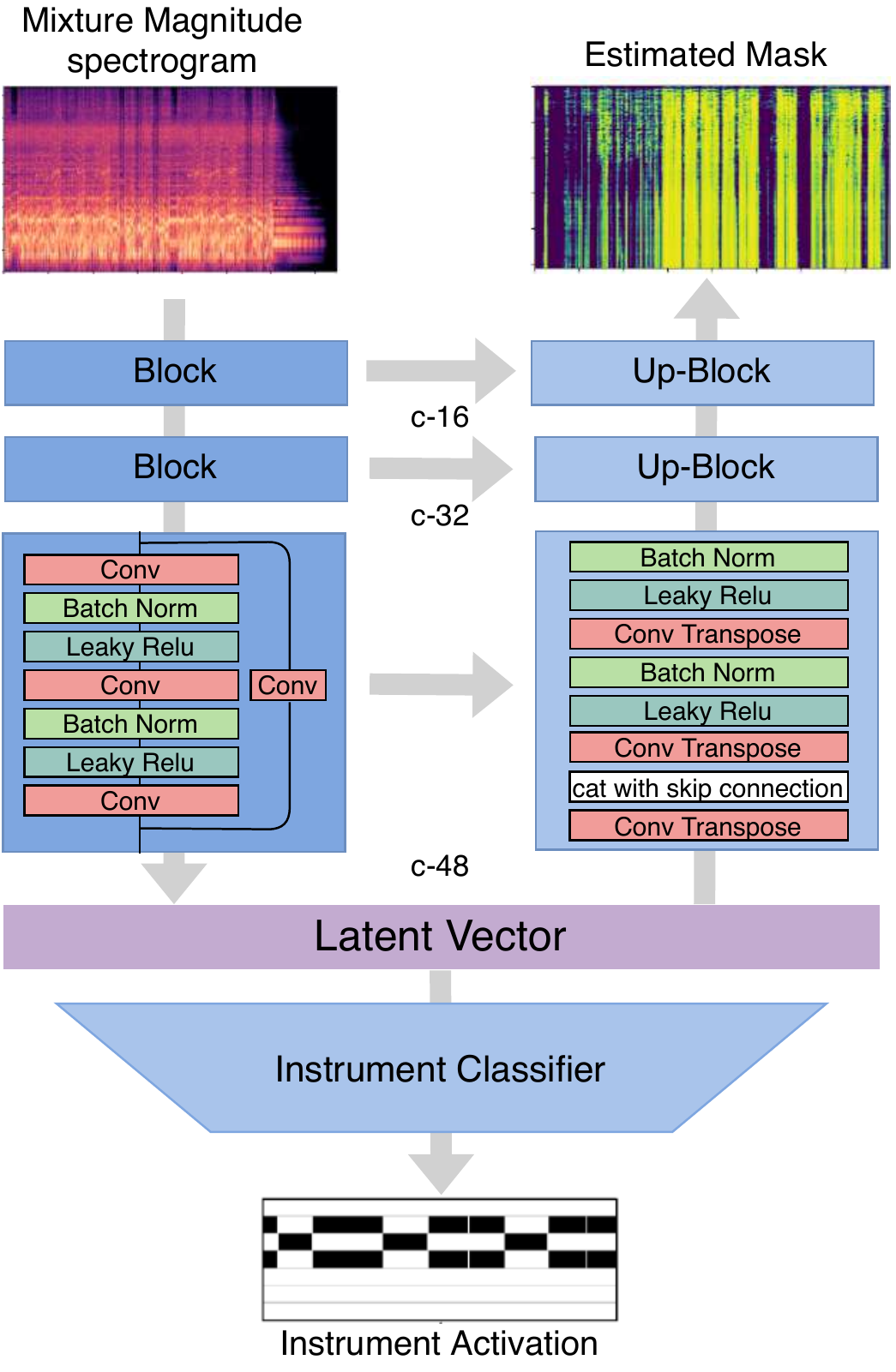}
    \caption{Multitask model structure for our proposed source separation system. Block is the residual block composed of convolutional layers while Up-Block is the residual block composed of transposed convolutional layers. ``c'' is the number of features.}
    \label{fig: model}
\end{figure}

\section{Method}
We propose a U-net-based \cite{ronneberger2015u} multitask structure to incorporate instrument recognition with music source separation, which we refer to as Instrument Aware Source Separation (IASS) system. An overview of the model is shown in \figref{fig: model}. 
Although the multitask approach shows similarities with previous approaches (compare \cite{manilow2019simultaneous}), we design our model with a different goal: instead of just learning a joint representation using the multitask structure, our model uses estimated labels from multitask learning during inference to improve source separation estimation. 
\subsection{Model structure}

The U-net structure has been found useful for image decomposition \cite{Ronneberger2015UNetCN}, a task with general similarities to source separation. The skip connections  of U-net  enable the model to learn from both high-level and low-level features leading to its success for music source separation \cite{jansson2017singing, stoller2018wave, spleeter2019}.

Our model differs from previous U-net-based source separation systems by using a residual block instead of a CNN in each layer. The residual block allows the information from the current layer to be fed into a layer 2 hops away and deepens the structure. Each encoder and decoder contains three blocks with each block containing three convolutional or transposed convolutional layers, respectively, two batch normalization layers, and two leaky ReLU layers. The multitask objective is achieved by attaching a CNN classifier to the latent vector. This classifier predicts the instrument activity and has four transposed convolutional layers and three batch normalization layers in between. The last convolutional and transposed convolutional layers in each block feature a (3,1) filter size for up-sampling or down-sampling, while the others have a (3,3) filter size. During training, we use a Mean Square Error (MSE) loss for source separation and a Binary Cross-Entropy (BCE) loss for the prediction of the instrument activity. A hyperparameter $\alpha$ is manually tuned to balance these two loss functions:
\begin{equation} 
L = L_\mathrm{MSE} + \alpha L_\mathrm{BCE} .
\label{eq: loss}
\end{equation}

\begin{figure}[t]
    \centering
    \includegraphics[width =
    \columnwidth]{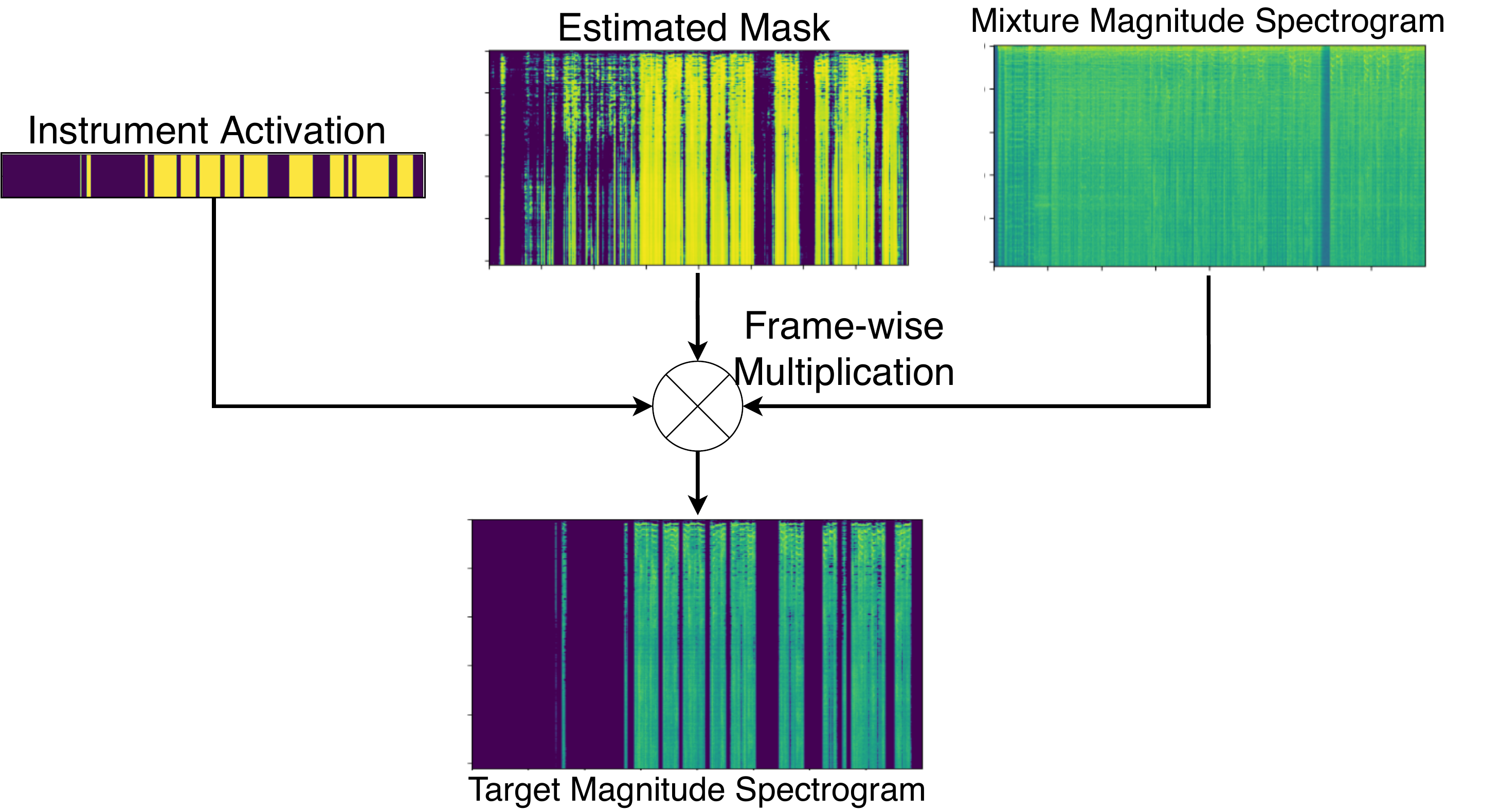}
    \caption{Using instrument activation as a weight to filter the estimated mask, which will be used to multiply with the mixture of the magnitude spectrogram.}
    \label{fig: filter}
\end{figure}

\begin{figure}[t]
    \centering
    \includegraphics[width =
    1\columnwidth]{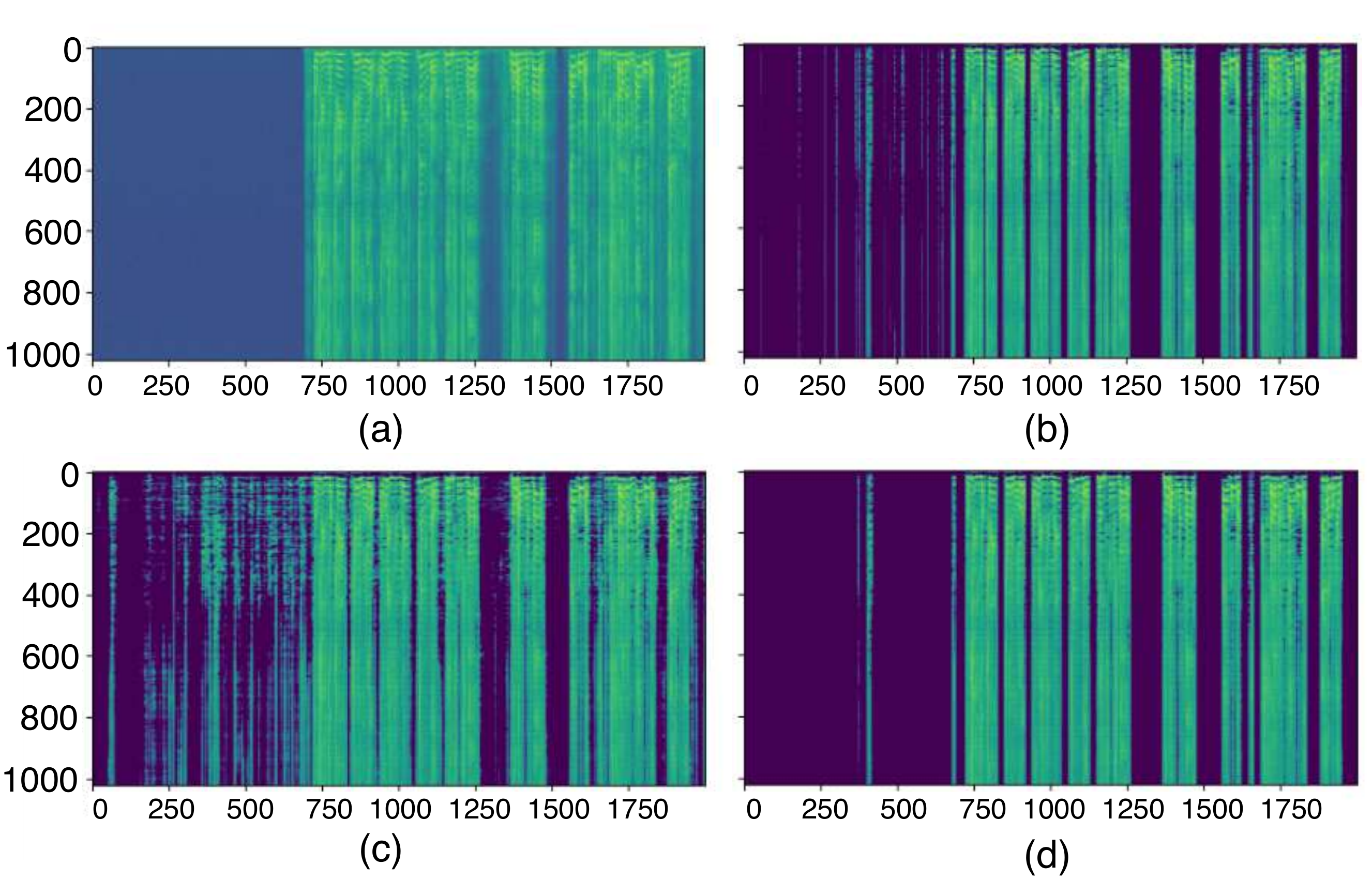}
    \caption{First 2000 frames of the separated vocal track from the song \textit{Angels In Amplifiers~---~I'm Alright} from the MUSDB-HQ dataset, visualizing different post-processing methods for applying instrument activity as a weight on the predicted magnitude spectrogram: (a) ground truth spectrogram, (b) predicted spectrogram without post-processing, (c) predicted spectrogram with raw predicted instrument labels as a weight, (d) predicted spectrogram with smoothed predicted instrument labels.}
\label{fig: output}
\end{figure}

After successful training, the predicted instrument activity is used as a binary weight to multiply with the magnitude spectrogram along the time dimension.
By doing so, the instrument labels are able to suppress the frames not containing any target instrument as shown in Figure~\ref{fig: filter}. However, this binary instrument mask has two potential problems. First, false negatives of the predicted labels might mistakenly suppress wanted components in the spectrogram. Figure \ref{fig: output} (b) exemplifies that around frames 800 and 1000 with gaps caused by false negative prediction within a continuous sound. Even if the gaps have only a the length of a few milliseconds, it will have negative impact on the perceived quality. Second, the binary mask might cause repeated abrupt switching between silence and sound, which might lead to artifacts such as musical noise further decreasing the perceived quality. To address these problems, a median filter is applied to smooth the predicted instrument activities. The influence is discussed in Sect.~\ref{sec:exp}. 

An implementation of our system is publicly available online.\footnote{\href{https://biboamy.github.io/Source_Separation_Inst/}{ https://biboamy.github.io/Source\textunderscore Separation\textunderscore Inst}}

\subsection{Data representation} \label{data_rep}
We extract magnitude spectrograms with a window length and hop size of 4096 and 1024 samples, respectively, at a sample rate of \unit[44100]{Hz} for the input of our source separation model. The same magnitude spectrogram  is  extracted for the target audio reference. 
The instrument activity ground truth is at the frame level, meaning there is a binary label for each instrument to show whether the instrument is active or not in each time frame. Instrument labels have the same time resolution as the input spectrogram. We use the original activation probability computed by both datasets \cite{bittner2014medleydb, gururani2017mixing}, and binarize the activation with a threshold of 0.5 as suggested. 

\section{Experiment}\label{sec:exp}

To show the efficiency of our proposed model, we first compare our model with the baseline Open-Unmix model  \cite{stoter2019open} on the MUSDB-HQ dataset. Note that we choose MUSDB-HQ instead of MUSDB because we want to obtain a high-quality separation system without potential coding artifacts. The audio of MUSDB is encoded in a lossy format while MUSDB-HQ provides the raw audio data. Other than that, there is no difference between the two datasets. However, since MUSDB-HQ is a newly released dataset it complicates comparing our results to other approaches directly as most of the previous systems have not been evaluated on the MUSDB-HQ dataset.
Both the baseline and the proposed method are then evaluated on the MM dataset with six different instruments. For each source, a separate model is trained for both the baseline and our proposed method. Finally, an ablation study is conducted to investigate the impact of instrument labels and median filter on the source separation results. We train our models with the Adam optimizer with a learning rate of 0.001 and apply early stopping if the validation loss does not change for 100 epochs. 

\begin{table}
\centering
\begin{tabular*}{\columnwidth}{l| @{\extracolsep{\fill}} cccc}
\textbf{Method}& \textbf{Vocals}& \textbf{Bass} & \textbf{Drums} & \textbf{Other}\\
\hline\hline
IASS 3 blocks&6.46&4.18&5.56&4.19\\
IASS 4 blocks&6.51&4.25&5.15&4.38
\end{tabular*}
\caption{SDR score for IASS source separation performance with 3 and 4 residual blocks.}
\label{tab:result_blocks}
\end{table}

\subsection{Dataset}\label{sec:data}
Two open-source datasets, MUSDB-HQ\cite{musdb18-hq} and the combination of Mixing Secrets \cite{gururani2017mixing} and MedleyDB\cite{bittner2014medleydb} dataset (MM dataset) are used for the experiments. MUSDB is the most widely used dataset for music source separation and contains four separated tracks: `Bass,' `Drums,' `Vocals,' and `Others.' The dataset has 150 full-length stereo music tracks. We use the data split proposed by St\"oter \cite{stoter2019open}: 86/14/50 songs for training, validation, and testing, respectively. Since data augmentation has been proven to be helpful \cite{takahashi2018mmdenselstm}, the following data augmentation is applied during the training. First, one track is randomly selected from each source and multiplied with a random gain factor ranging from 0.25 to 1.25. The starting point of each track is randomly chosen for chunking into a clip with length of \unit[6]{s}. Finally, the chunked audio clips from each source are remixed for training. Since the original MUSDB-HQ dataset does not include instrument activation labels, we apply the energy tracking method proposed for MedleyDB \cite{bittner2014medleydb} with a threshold of 0.5 to obtain the frame-level binary instrument activity labels.

The MM dataset contains 585 pieces of songs (330 from MedleyDB and 258 from Mixing Secrets) with more than 100 instruments and their individual tracks. We use the training and testing split proposed by Gururani et al.\ \cite{gururani2018instrument} for training (488 songs) and evaluating (100 songs) our system. 
The most frequently occurring 6 instruments are picked as target instruments: `Bass,' `Drums,' `Vocals,' `Electrical Guitar,' `Acoustic Guitar,' and `Piano.' One of the problems with this dataset is that not all of the songs provide parameters on how to remix the individual tracks into the mixture. Therefore, the volume of each track is adjusted to the same loudness (RMS) during training before applying the random gain as detailed above. In addition, each track is downmixed to a single channel. Furthermore, the data augmentation technique introduced above is applied to generate a large number of training samples from the MM dataset. We construct two separate groups of songs. One contains all the tracks including the target instrument, while another contains the tracks without the target instrument (``accompaniments''). There are total of 128 target tracks for `Acoustic Guitar,' 189 for `Piano,' 325 for `Electrical Guitar,' 374 for `Vocals,' 468 for `Drums,' and 458 for `Bass.' During training, we randomly select 1 to 5 tracks in the accompaniment pool to mix with the target instrument. By doing so, we can generate various combinations of training ``songs.'' The random chunking approach applied to MUSDB-HQ is also applied on MM dataset. The testing set also balances the loudness of each track. We filter out the songs from the testing set which do not contain any of the 6 target instruments, resulting in 20 songs with `Piano,' 23 songs with `Acoustic Guitar,' 54 songs with `Electrical Guitar,' 71 songs with `Vocals,' 76 songs with `Bass,' and 81 songs with `Drums.'

\subsection{Evaluation}
To reconstruct the waveforms from the resulting magnitude spectrograms, we multiply the magnitude spectrogram with the phase of the original complex spectrograms and apply the inverse short-time Fourier transform on the complex spectrogram. We do not use any post-processing such as Wiener filtering here to focus on the raw result without potentially confounding quality gains in post-processing. Therefore, the Open-Unmix post-processing is disabled for the evaluation.

The quality of the source separation is evaluated with the four most frequently used objective metrics: source to distortion ratio (SDR), source to interference ratio (SIR), source to artifact ratio (SAR), and Image to Spatial distortion Ratio (ISR) \cite{Fvotte2005BSS}. We use the \textit{museval} package for calculating the evaluation metrics \cite{SiSEC18}.

\begin{table}
\centering
\begin{tabular*}{\columnwidth}{l|l| @{\extracolsep{\fill}} cccc}
&\textbf{Method}&\textbf{SDR}& \textbf{SIR} &\textbf{SAR} &\textbf{ISR}\\
\hline\hline
Vocals&Open-Unmix&6.11&13.21&6.75&12.43\\
&IASS &\textbf{6.46}&\textbf{14.70}&\textbf{6.98}&\textbf{14.30}\\\hline
Bass&Open-Unmix&\textbf{4.48}&\textbf{8.23}&\textbf{5.40}&\textbf{10.29}\\
&IASS &4.18&7.30&4.52&6.85\\\hline
Drums&Open-Unmix&5.02&10.17&6.05&10.55\\
&IASS  &\textbf{5.56}&\textbf{10.74}&\textbf{6.86}&\textbf{10.92}\\\hline
Other&Open-Unmix&\textbf{4.23}&\textbf{9.90}&3.88&7.34\\
&IASS  &4.19&8.78&\textbf{4.70}&\textbf{9.32}\\
\end{tabular*}
\caption{BSS metrics for Open-Unmix and IASS on the MUSDB-HQ dataset.}
\label{tab:result_MUSDB-HQ}
\end{table}

\subsubsection{Source separation on MUSDB-HQ}

To allow for comparison with other systems trained on the MUSDB, the first experiment reports the result on MUSDB-HQ. 

In a first preliminary experiment we investigate whether adding more residual blocks influences the performance of the proposed method. We report the SDR score for the four sources of the MUSDB-HQ dataset in Table~\ref{tab:result_blocks}. The performance with three residual blocks and with four residual blocks is comparable on all instruments. For training efficiency, we use three residual blocks in the following experiment. 

Table~\ref{tab:result_MUSDB-HQ} shows the results of our proposed model compared to Open-Unmix. 
Our model outperforms the Open-Unmix model on `Vocals' and `Drums', performs equally on `Other', and slightly worse on `Bass'. This might be because `Bass' most likely to appear throughout the songs. As a result, the improvement of using the instrument activation weight is limited. The imbalanced activity might also impact the instrument classifier.



\subsubsection{Source separation on MM}

The results for the MM dataset are summarized in Table~\ref{tab:result_openunmix}. We re-trained the Open-Unmix model on the MM dataset by using the default training setting provided with their code. The results for the Ideal Binary Mask (IBM) (source code \cite{SiSEC18}) represent the best case scenario. The worst case scenario is represented by the results for input-SDR, which is the SDR score when using mixture as the input. Compared to MUSDB-HQ, the MM dataset has a larger amount of training data. It can be observed from Table~\ref{tab:result_openunmix} that our proposed model generally achieves better source separation performance on six instruments. We can also observe a trend that both models have higher scores on `Drums,' `Bass,' and `Vocals' than on `Electrical Guitar,' `Piano,' and `Acoustic Guitar.' 
This might be attributed to the fact that `Guitar,' `Piano,' and `Acoustic Guitar' have fewer training samples (cf.\ Sect.~\ref{sec:data}). Another possible reason is that the more complicated spectral structure of polyphonic instruments such as `Guitar' and `Piano' make the separation task more challenging. 

\begin{table}
\centering
\begin{tabular}{l|>{\centering}p{0.07\textwidth}>{\centering}p{0.07\textwidth}>{\centering}p{0.07\textwidth}>{\centering\arraybackslash}p{0.07\textwidth}}
&\textbf{Open-Unmix}&\textbf{IASS}& \textbf{IBM}&\textbf{input-SDR}\\
\hline
\hline
Vocals&3.68&4.78&6.49&-6.24\\\hline
Elecgtr&1.55&1.77&4.56&-5.90\\\hline
Acgtr&0.95&1.29&3.38&-6.65\\\hline
Piano&1.08&1.91&3.63&-6.31\\\hline
Bass&4.04&5.26&5.34&-5.77\\\hline
Drums&4.45&4.89&6.23&-6.05\\
\end{tabular}
\caption{SDR score for Open-Unmix, IASS, an ideal binary mask and input-SDR.}
\label{tab:result_openunmix}
\end{table}

\subsubsection{Instrument activity detection}

\begin{figure}[t]
    \centering
    \includegraphics[width =
    \columnwidth]{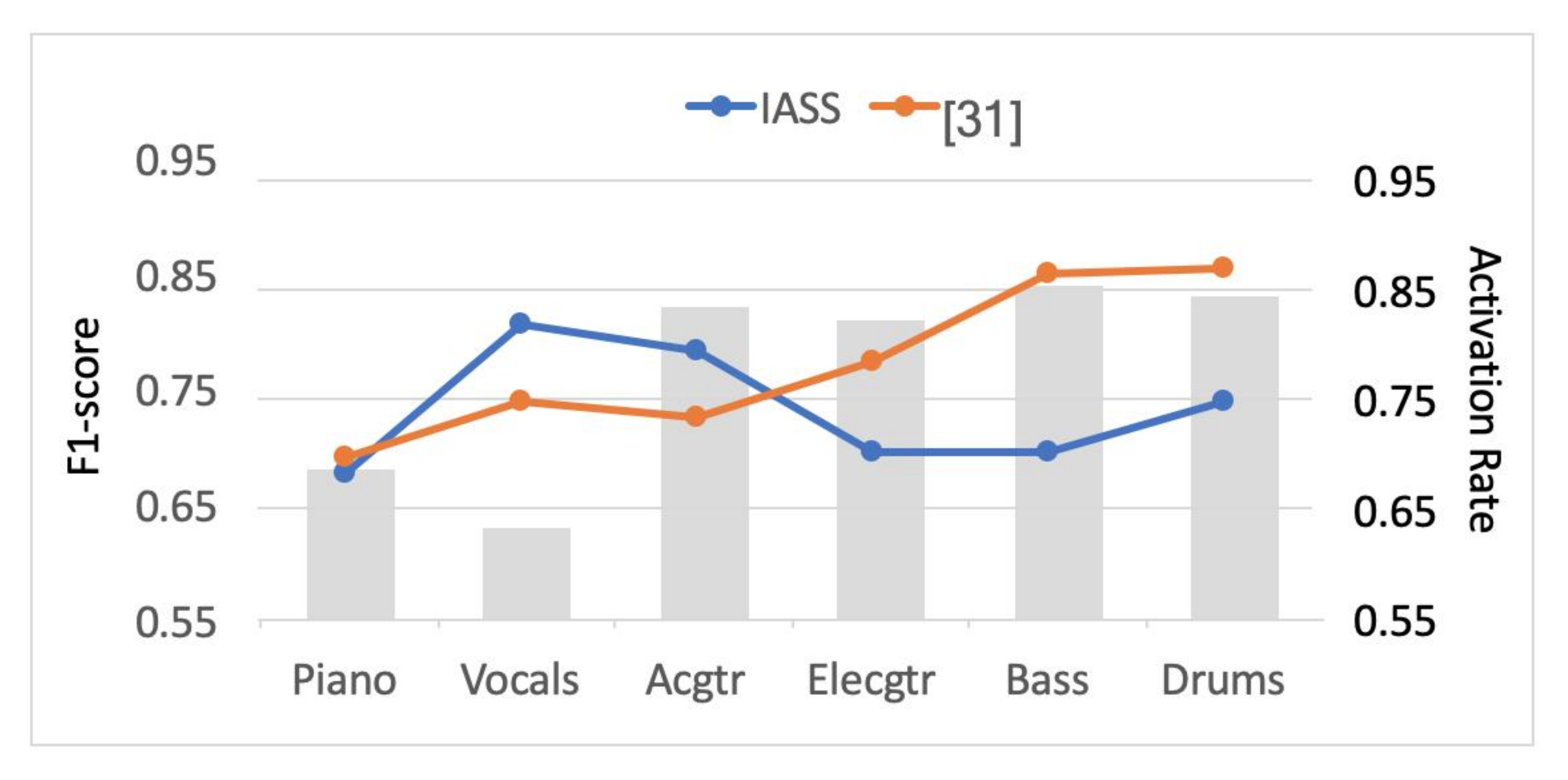}
    \caption{IAD result for both Gururani et al.'s method (orange) and our IASS (blue) with label aggregation. Indicated in gray is the activation rate (percentage of the training frames containing positive activity labels).}
    \label{fig: inst_result}
\end{figure}

While our system's source separation performance was the primary concern, the accuracy of the instrument predictions is also of interest. Our classifier output is compared to the model proposed by Gururani et al.\  \cite{gururani2018instrument}, which was trained and evaluated on the same MM dataset. Note that this comparison is still not completely valid as their system uses multi-label prediction while our model is single-label. Still, it can provide some insights into how well our system predicts the instrument activity. As Gururani et al.'s system predicts instrument labels with a time resolution of \unit[1]{s}, the output resolution of our prediction has to be reduced. For each second, all the estimated activations are aggregated by calculating their median. 
Furthermore, instrument subcategories from their work are combined. For example, female and male singers are combined into `Vocals,'  electrical bass and double bass are combined into `Bass,' electrical and acoustic piano are combined into `Piano' and clean and distorted electrical guitar are combined into `Electrical Guitar.' We report the AUC score in \figref{fig: inst_result}. 

We can make the following observations. First, `Piano,' `Electrical Guitar,' and `Bass' tend to have lower detection rates. This might be because all these instrument categories include both acoustic and electric instruments which the model might easily confuse with the background music. 
This might also influence the source separation performance. The result can explain that in Table~\ref{tab:result_inst}, `Vocals' have the highest increase in the average score when applying instrument labels since `Vocals' has better instrument detection accuracy. In contrast, `Bass' has a lower increase since it has poorer instrument detection results. Second, from \figref{fig: inst_result} we can observe that `Vocals' and `Piano' have a lower activation rate, which means the model has fewer sound samples containing `Vocals' and `Piano' during training. This aligns with the highest SIR score increase on `Vocals' and `Piano' in Table~\ref{tab:result_inst} when instrument activation is added, since instrument activation can help suppress the interference at the non-active frames. This also shows the potential of our model to be used on instruments which only appear in the song infrequently. 

\begin{table}
\centering
\begin{tabular*}{\columnwidth}{l @{\extracolsep{\fill}} |l|c|cccc}
&\textbf{Train} &\textbf{Test} &\textbf{SDR}& \textbf{SIR} &\textbf{SAR} &\textbf{Avg}\\
\hline\hline
Vocals&\xmark&\xmark&4.26&8.58&4.48&5.77\\
&\cmark&\xmark&3.94&8.48&4.69&5.70\\
&\cmark&\cmark&\textbf{4.78}&\textbf{11.62}&\textbf{5.31}&\textbf{7.24}\\\hline
Elecgtr&\xmark&\xmark&1.75&0.61&\textbf{4.94}&\textbf{2.46}\\
&\cmark&\xmark&\textbf{1.82}&1.27&4.29&2.43\\
&\cmark&\cmark&1.77&\textbf{1.64}&4.48&2.63\\\hline
Acgtr&\xmark&\xmark&1.11&0.75&2.42&1.43\\
&\cmark&\xmark&1.15&0.48&\textbf{2.52}&1.38\\
&\cmark&\cmark&\textbf{1.29}&\textbf{1.80}&2.45&\textbf{1.85}\\\hline
Piano&\xmark&\xmark&1.55&2.97&2.13&2.31\\
&\cmark&\xmark&1.70&3.16&2.06&2.22\\
&\cmark&\cmark&\textbf{1.91}&\textbf{4.17}&\textbf{2.15}&\textbf{2.74}\\\hline
Bass&\xmark&\xmark&4.10&\textbf{8.34}&4.74&5.72\\
&\cmark&\xmark&4.12&7.82&\textbf{5.10}&5.68\\
&\cmark&\cmark&\textbf{4.34}&{8.13}&\textbf{5.10}&\textbf{5.85}\\\hline
Drums&\xmark&\xmark&4.50&9.54&5.15&6.40\\
&\cmark&\xmark&4.38&9.87&4.95&6.40\\
&\cmark&\cmark&\textbf{4.89}&\textbf{10.72}&\textbf{5.26}&\textbf{6.96}
\end{tabular*}
\caption{Ablation study for IASS source separation performance training and evaluating with (\cmark) or without (\xmark) instrument labels.}
\label{tab:result_inst}
\end{table}

\subsection{Ablation study}
In this experiment, we investigate the impact of the instrument labels on our model. 
First, the IASS model is trained and evaluated without using instrument labels, i.e., as a standard U-net without instrument classifier on the latent vector. The model will only be updated by the MSE between the ground-truth magnitude spectrogram and predicted spectrogram ($\alpha = 0$). For testing, all instrument ``predictions'' are set to $1$. 
Second, the IASS model is trained with instrument labels but evaluated without instrument labels. This is a traditional multitask scenario: the model will be trained with both the MSE and the BCE losses in Eq.~(\ref{eq: loss}). However, during evaluation, the output magnitude spectrogram is not weighted by the instrument activity  (predictions equal $1$). 
Third, we include the IASS results from Table~\ref{tab:result_openunmix}~---computed with both losses and using the instrument predictions as mask weights---~for convenience.

The results are shown in Table~\ref{tab:result_inst}. It can be observed that using instrument labels as a weight generally leads to a better performance than without using instrument labels. The result also somewhat unexpectedly shows that training with instrument detection loss influences source separation performance, as the average quality score is often lower when training with the multitask loss. One possible reason for this is that adding the IAD sub-task forces more information to be passed to the bottom layers, where the resolution is compressed. We argue, however, that the multitask learning structure does bring an extra benefit to the system: using the instrument activity predictions as a weight leads to better separation quality. Figure~\ref{fig: output} visualizes the effect on one of the songs from the MUSDB-HQ dataset. This song does not have any vocals before \unit[16]{s}, which is around time frame $700$. Subfigure (a) shows the ground truth magnitude spectrogram before applying instrument labels while (c) and (d) show the predicted spectrograms after applying the instrument activations or the smoothed instrument activations, respectively. Both the false positive predictions in the beginning before time frame $700$ as well as the false negative predictions around frames $800$ and $1000$ have been repaired by using smoothed activations. This result is consistent with the results in Table~\ref{tab:result_inst} where SIR has the highest increase: interferences are more successfully suppressed. 

\begin{figure}[t]
    \centering
    \includegraphics[width =
    \columnwidth]{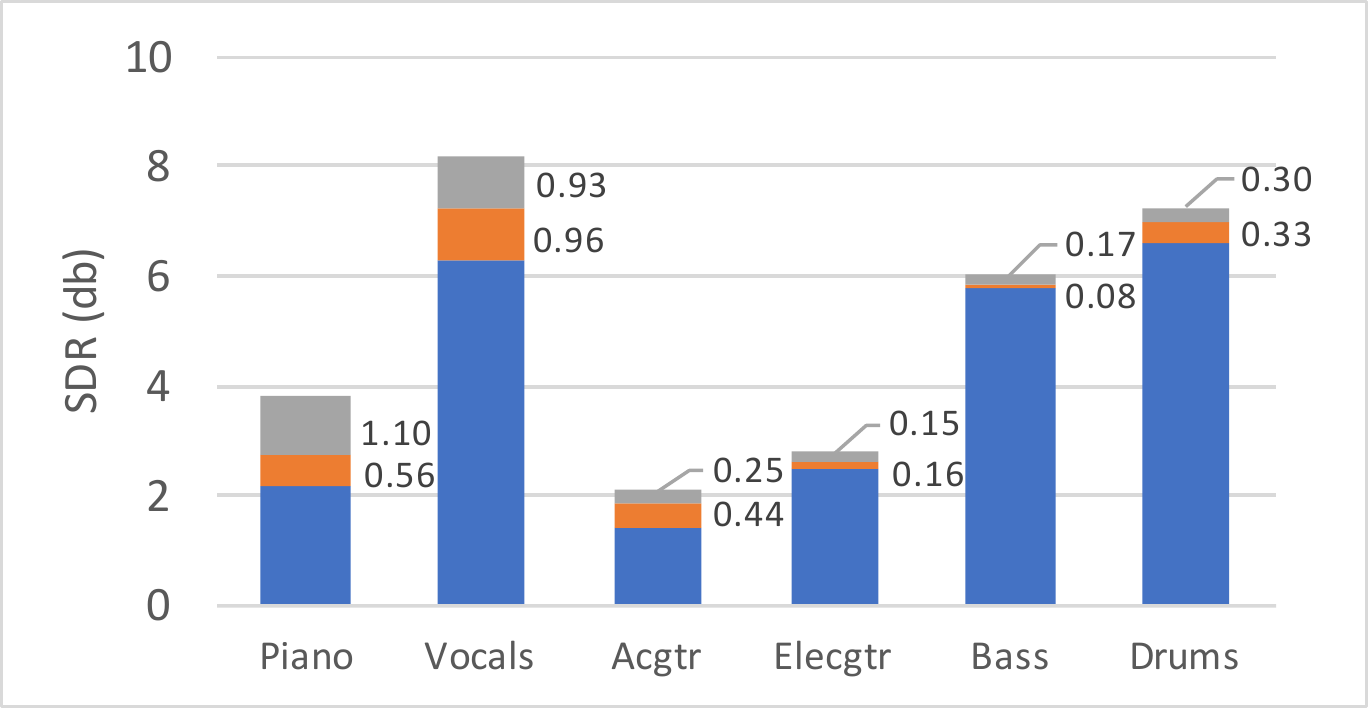}
    \caption{Ablation study for IASS source separation performance with or without median filter on instrument labels. The orange bar shows the increase of the average score (average of SDR, SIR, and SAR) after applying median filtering. Gray shows the average score increase when using instrument ground truth labels instead of estimated labels.
    }
    \label{fig: filter_gt}
\end{figure}

Furthermore, we investigate the influence of median filtering the predicted instrument activity on our results. Figure~\ref{fig: filter_gt} shows the performance of our proposed source separation model with and without applying the median filter on the predicted instrument activities. Using the median filter generally increases performance across all instruments as it eliminates spurious prediction errors. 

Finally we are using the oracle ground truth labels instead of the estimated labels as the weight. As we can observe from Figure~\ref{fig: filter_gt}, using the ground truth labels brings an average score increase in all instruments, especially for vocals and piano. This can be seen as the upper-bound best case scenario of our instrument-activity-weighted model and emphasizes the potential for improvement when combining instrument prediction with source separation.

\section{Conclusion}
In this paper, we proposed a novel multitask structure combining instrument activation detection with multi-instrument source separation. We utilize a large dataset to evaluate on various instruments and show that our model achieves equal or better separation quality than the baseline Open-Unmix model. The ablation study also shows that using instrument activation as a weight is able to correct the false estimation from the source separation task and improve source separation performance. In summary, the main  contributions of this work are the proposal of a multitask learning structure combining IAD with source separation, and insights into using new open-source datasets to increase the number of separable instrument categories.   

We have identified several directions for future extensions of this model. First, we plan to increase the number of target instruments
by combining synthesized data with the MM dataset especially for underrepresented instrument classes. Second, we plan to incorporate other tasks, such as multi-pitch estimation, into our current multi-task structure \cite{Nakano2019JointSP}.
Third, we will explore using multi-label instrument detection to separate multiple instruments at the same time. Lastly, we will explore post-processing methods such as Wiener filter to improve our system's quality. 


\section{Acknowledgment}\label{sec:acknowledgement}
We gratefully acknowledge NVIDIA Corporation (Santa Clara, CA, United States) who supported this research by providing a Titan X GPU via the NVIDIA GPU Grant program.

\bibliography{ISMIRtemplate}

\begin{thebibliography}{10}

\bibitem{adali2014source}
Tülay Adali, Christian Jutten, Arie Yeredor, Andrzej Cichocki, and Eric
  Moreau.
\newblock From raw audio to a seamless mix: creating an automated {DJ} system
  for {Drum} and {Bass}.
\newblock {\em IEEE Signal Processing Magazine}, pages 16--17, 2014.

\bibitem{Bittner2018MultitaskLF}
Rachel~M. Bittner, Brian McFee, and Juan~Pablo Bello.
\newblock Multitask learning for fundamental frequency estimation in music.
\newblock {\em arXiv preprint arXiv:1809.00381}, 2018.

\bibitem{bittner2014medleydb}
Rachel~M. Bittner, Justin Salamon, Mike Tierney, Matthias Mauch, Chris Cannam,
  and Juan~Pablo Bello.
\newblock Medleydb: A multitrack dataset for annotation-intensive mir research.
\newblock In {\em Proceedings of the International Society for Music
  Information Retrieval Conference (ISMIR)}, pages 155--160, 2014.

\bibitem{Bck2019MultiTaskLO}
Sebastian B{\"o}ck, Matthew E.~P. Davies, and Peter Knees.
\newblock Multi-task learning of tempo and beat: Learning one to improve the
  other.
\newblock In {\em Proceedings of the International Society for Music
  Information Retrieval Conference (ISMIR)}, pages 569--576, 2019.

\bibitem{Chan2015VocalAI}
Tak-Shing Chan, Tzu-Chun Yeh, Zhe-Cheng Fan, Hung-Wei Chen, Li~Su, Yi-Hsuan
  Yang, and Jyh-Shing~Roger Jang.
\newblock Vocal activity informed singing voice separation with the ikala
  dataset.
\newblock In {\em IEEE International Conference on Acoustics, Speech and Signal
  Processing (ICASSP)}, pages 718--722, 2015.

\bibitem{defossez2019music}
Alexandre D{\'e}fossez, Nicolas Usunier, L{\'e}on Bottou, and Francis Bach.
\newblock Music source separation in the waveform domain.
\newblock {\em arXiv preprint arXiv:1911.13254}, 2019.

\bibitem{Fvotte2005BSS}
C{\'e}dric F{\'e}votte, R{\'e}mi Gribonval, and Emmanuel Vincent.
\newblock Bss\textunderscore eval toolbox user guide -- revision 2.0.
\newblock In {\em IRISA Technical Report 1706, Rennes, France}, 2005.

\bibitem{gururani2017mixing}
Siddharth Gururani and Alexander Lerch.
\newblock Mixing secrets: a multi-track dataset for instrument recognition in
  polyphonic music.
\newblock In {\em Proceedings of the International Society for Music
  Information Retrieval Conference (ISMIR)}, 2017.
\newblock Late-breaking paper.

\bibitem{gururani2018instrument}
Siddharth Gururani, Cameron Summers, and Alexander Lerch.
\newblock Instrument activity detection in polyphonic music using deep neural
  networks.
\newblock In {\em Proceedings of the International Society for Music
  Information Retrieval Conference (ISMIR)}, pages 569--576, 2018.

\bibitem{spleeter2019}
Romain Hennequin, Anis Khlif, Felix Voituret, and Manuel Moussallam.
\newblock Spleeter: A fast and state-of-the art music source separation tool
  with pre-trained models.
\newblock In {\em Proceedings of the International Society for Music
  Information Retrieval Conference (ISMIR)}, 2018.
\newblock Late-breaking paper.

\bibitem{Hsu2010OnTI}
Chao-Ling Hsu and Jyh-Shing~Roger Jang.
\newblock On the improvement of singing voice separation for monaural
  recordings using the mir-1k dataset.
\newblock {\em IEEE Transactions on Audio, Speech, and Language Processing},
  18:310--319, 2010.

\bibitem{Hung2019MultitaskLF}
Yun-Ning Hung, Yi-An Chen, and Yi-Hsuan Yang.
\newblock Multitask learning for frame-level instrument recognition.
\newblock In {\em IEEE International Conference on Acoustics, Speech and Signal
  Processing (ICASSP)}, pages 381--385, 2019.

\bibitem{Hung2018FramelevelIR}
Yun-Ning Hung and Yi-Hsuan Yang.
\newblock Frame-level instrument recognition by timbre and pitch.
\newblock In {\em Proceedings of the International Society for Music
  Information Retrieval Conference (ISMIR)}, pages 569--576, 2018.

\bibitem{jansson2017singing}
Andreas Jansson, Eric Humphrey, Nicola Montecchio, Rachel Bittner, Aparna
  Kumar, and Tillman Weyde.
\newblock Singing voice separation with deep {U-Net} convolutional networks.
\newblock In {\em Proceedings of the International Society for Music
  Information Retrieval Conference (ISMIR)}, page 323–332, 2017.

\bibitem{lee2019audio}
Jie~Hwan Lee, Hyeong-Seok Choi, and Kyogu Lee.
\newblock Audio query-based music source separation.
\newblock In {\em Proceedings of the International Society for Music
  Information Retrieval Conference (ISMIR)}, pages 878--885, 2019.

\bibitem{liu2019dilated}
Jen-Yu Liu and Yi-Hsuan Yang.
\newblock Dilated convolution with dilated gru for music source separation.
\newblock In {\em International Joint Conferences on Artificial Intelligence},
  pages 4718--4724, 2019.

\bibitem{SiSEC16}
Antoine Liutkus, Fabian-Robert St{\"o}ter, Zafar Rafii, Daichi Kitamura,
  Bertrand Rivet, Nobutaka Ito, Nobutaka Ono, and Julie Fontecave.
\newblock The 2016 signal separation evaluation campaign.
\newblock In {\em International conference on latent variable analysis and
  signal separation}, pages 323--332, 2017.

\bibitem{manilow2019simultaneous}
Ethan Manilow, Prem Seetharaman, and Bryan Pardo.
\newblock Simultaneous separation and transcription of mixtures with multiple
  polyphonic and percussive instruments.
\newblock In {\em IEEE International Conference on Acoustics, Speech and Signal
  Processing (ICASSP)}, pages 771--775, 2020.

\bibitem{Miron2017MonauralSS}
Marius Miron, Jordi Janer, and Emilia G{\'o}mez.
\newblock Monaural score-informed source separation for classical music using
  convolutional neural networks.
\newblock In {\em Proceedings of the International Society for Music
  Information Retrieval Conference (ISMIR)}, pages 569--576, 2017.

\bibitem{Nakano2019JointSP}
Tomoyasu Nakano, Kazuyoshi Yoshii, Yiming Wu, Ryo Nishikimi, Kin Wah~Edward
  Lin, and Masataka Goto.
\newblock Joint singing pitch estimation and voice separation based on a neural
  harmonic structure renderer.
\newblock {\em 2019 IEEE Workshop on Applications of Signal Processing to Audio
  and Acoustics (WASPAA)}, pages 160--164, 2019.

\bibitem{nugraha2016multichannel}
Aditya~Arie Nugraha, Antoine Liutkus, and Emmanuel Vincent.
\newblock Multichannel music separation with deep neural networks.
\newblock In {\em European Signal Processing Conference (EUSIPCO)}, pages
  1748--1752, 2016.

\bibitem{paulus2005drum}
Jouni Paulus and Tuomas Virtanen.
\newblock Drum transcription with non-negative spectrogram factorisation.
\newblock In {\em European Signal Processing Conference}, pages 1--4, 2005.

\bibitem{musdb18}
Zafar Rafii, Antoine Liutkus, Fabian-Robert St{\"o}ter, Stylianos~Ioannis
  Mimilakis, and Rachel Bittner.
\newblock The {MUSDB18} corpus for music separation, December 2017.
\newblock [Online] \url{https://doi.org/10.5281/zenodo.1117372}.

\bibitem{rafii2018overview}
Zafar Rafii, Antoine Liutkus, Fabian-Robert Stoter, Stylianos~Ioannis
  Mimilakis, Derry FitzGerald, and Bryan Pardo.
\newblock An overview of lead and accompaniment separation in music.
\newblock {\em IEEE/ACM Transactions on Audio, Speech and Language Processing
  (TASLP)}, 26(8):1307--1335, 2018.

\bibitem{musdb18-hq}
Zafar Rafii, Antoine Liutkus, Fabian-Robert Stöter, Stylianos~Ioannis
  Mimilakis, and Rachel Bittner.
\newblock {MUSDB18-HQ} - an uncompressed version of {MUSDB18}, August 2019.

\bibitem{Ronneberger2015UNetCN}
Olaf Ronneberger, Philipp Fischer, and Thomas Brox.
\newblock U-net: Convolutional networks for biomedical image segmentation.
\newblock In {\em International Conference on Medical Image Computing and
  Computer Assisted Intervention}, 2015.

\bibitem{ronneberger2015u}
Olaf Ronneberger, Philipp Fischer, and Thomas Brox.
\newblock U-net: Convolutional networks for biomedical image segmentation.
\newblock In {\em International Conference on Medical image computing and
  computer-assisted intervention}, pages 234--241, 2015.

\bibitem{seetharaman2019class}
Prem Seetharaman, Gordon Wichern, Shrikant Venkataramani, and Jonathan Le~Roux.
\newblock Class-conditional embeddings for music source separation.
\newblock In {\em IEEE International Conference on Acoustics, Speech and Signal
  Processing (ICASSP)}, pages 301--305, 2019.

\bibitem{slizovskaia2019end}
Olga Slizovskaia, Leo Kim, Gloria Haro, and Emilia Gomez.
\newblock End-to-end sound source separation conditioned on instrument labels.
\newblock In {\em IEEE International Conference on Acoustics, Speech and Signal
  Processing (ICASSP)}, pages 306--310, 2019.

\bibitem{stoller2018wave}
Daniel Stoller, Sebastian Ewert, and Simon Dixon.
\newblock Wave-u-net: A multi-scale neural network for end-to-end audio source
  separation.
\newblock {\em arXiv preprint arXiv:1806.03185}, 2018.

\bibitem{SiSEC18}
Fabian-Robert St{\"o}ter, Antoine Liutkus, and Nobutaka Ito.
\newblock The 2018 signal separation evaluation campaign.
\newblock In {\em International Conference on Latent Variable Analysis and
  Signal Separation}, pages 293--305, 2018.

\bibitem{stoter2019open}
Fabian-Robert St{\"o}ter, Stefan Uhlich, Antoine Liutkus, and Yuki Mitsufuji.
\newblock Open-unmix-a reference implementation for music source separation.
\newblock {\em Journal of Open Source Software}, 2019.

\bibitem{Swaminathan2019ImprovingSV}
Rupak Swaminathan and Alexander Lerch.
\newblock Improving singing voice separation using attribute-aware deep
  network.
\newblock {\em International Workshop on Multilayer Music Representation and
  Processing (MMRP)}, pages 60--65, 2019.

\bibitem{takahashi2018mmdenselstm}
Naoya Takahashi, Nabarun Goswami, and Yuki Mitsufuji.
\newblock Mmdenselstm: An efficient combination of convolutional and recurrent
  neural networks for audio source separation.
\newblock In {\em International Workshop on Acoustic Signal Enhancement
  (IWAENC)}, pages 106--110, 2018.

\bibitem{Uhlich2015DeepNN}
Stefan Uhlich, Franck Giron, and Yuki Mitsufuji.
\newblock Deep neural network based instrument extraction from music.
\newblock In {\em IEEE International Conference on Acoustics, Speech and Signal
  Processing (ICASSP)}, pages 2135--2139, 2015.

\bibitem{uhlich2017improving}
Stefan Uhlich, Marcello Porcu, Franck Giron, Michael Enenkl, Thomas Kemp, Naoya
  Takahashi, and Yuki Mitsufuji.
\newblock Improving music source separation based on deep neural networks
  through data augmentation and network blending.
\newblock In {\em IEEE International Conference on Acoustics, Speech and Signal
  Processing (ICASSP)}, pages 261--265, 2017.

\bibitem{veire2018raw}
Len~Vande Veire and Tijl De~Bie.
\newblock From raw audio to a seamless mix: creating an automated dj system for
  drum and bass.
\newblock {\em EURASIP Journal on Audio, Speech, and Music Processing},
  2018(1):13, 2018.

\end{thebibliography}

\end{document}